\ifx\pdftexversion\undefined
  \documentclass[preprint,dvips]{aastex}
\else
  \documentclass[preprint,pdftex]{aastex}
\fi

\newcommand{\be}{\begin{equation}}
\newcommand{\ee}{\end{equation}}
\newcommand{\bt}{\begin{table} \begin{center}}
\newcommand{\et}{\end{center} \end{table}}
\newcommand{\ba}{\begin{eqnarray}}
\newcommand{\ea}{\end{eqnarray}}

\newcommand{\eg}{{\it e.g.~}}
\newcommand{\cf}{{\it c.f.~}}
\def\d{{\rm d}}

\def\dd#1#2{\frac{\d #1}{\d #2}}

\def\eqref#1{Equation~(\ref{eq:#1})}

\begin{document}

\newcommand{\bfi}{{\bf B}} \newcommand{\efi}{{\bf E}}
\newcommand{\lel}{{\lambda_e^{\!\!\!\!-}}}
\newcommand{\me}{m_e}
\newcommand{\mcs}{{m_e c^2}}
\def\ho{{\hat {\bf o}}}
\def\hm{{\hat {\bf m}}}
\def\hx{{\hat {\bf x}}}
\def\hy{{\hat {\bf y}}}
\def\hz{{\hat {\bf z}}}
\def\hom{{\hat{\mathbf{\omega}}}}
\def\hr{{\hat {\bf r}}}
\def\omv{\mathbf{\omega}}

\title{R-Modes on Rapidly Rotating, Relativistic Stars : II. Blackbody Emission}%
\author{Jeremy S. Heyl\altaffilmark{1,2}}
\altaffiltext{1}{Chandra Fellow; Harvard-Smithsonian Center for
  Astrophysics, MS-51, 
60 Garden Street, Cambridge MA 02138, United States}
\altaffiltext{2} 
{Current Address: Department of Physics and Astronomy; 
University of British Columbia; Vancouver, BC V6T 1Z1, Canada; 
heyl@physics.ubc.ca}

\begin{abstract}
Rossby waves (or r-modes) on the surface of a neutron star have become
a leading model for the oscillations observed during the tail of
Type-I X-ray bursts.  Their frequency evolution matches well with the
observed frequency drifts of the oscillations in the bursts, and the
burning appears to excite these waves quite naturally.  This paper
addresses the detailed shape of the expected flux profiles from
r-modes on neutron stars as a function of energy.  R-modes naturally
account for both the small amplitude of the observed oscillations and
their lack of harmonic content.  However, the model predicts that the
oscillation at higher energies leads the lower energy variation.  The
observed oscillations have the opposite trend which possibly indicates
that the higher energy photons are upscattered in the plasma
surrounding the neutron star and therefore delayed.
\end{abstract}

\section{Introduction}
\label{sec:introduction}

\citet{Heyl01typei} proposed that the oscillations in the tail of
Type-I X-ray bursts are the signatures of Rossby waves travelling on
the surface of the neutron star.  This model predicts a specific
pattern of hot and cold regions on the surface of the star.
\citet{1999ApJ...519L..73F} proposed that spectral dependence of
Type-I X-ray burst oscillations may provide an important probe of the
mass and radius of neutron stars.
\citet{2002ApJ...581..550M,2003ApJ...595.1066M} examined the observed
energy dependence and harmonic content of burst oscillations.  
\citet{2005ApJ...618..451C} have studied the propagation of radiation
in the equatorial plane surrounding rotating neutron stars with
realistic equations of state in a full general relativistic
treatment with a hot-spot model for the emission.  Leaving the
equatorial plane in a general spacetime makes the treatment of photon
propagation much more complicated. 

This paper fills a gap in these various discussions.  Specifically the
emission from the surface of the star takes the form of a r-mode
\citep{Heyl01typei} that can account for the observed frequency shifts of
Type-I x-ray burst oscillations from rapidly rotating neutron stars in
low-mass x-ray binaries and combines this emission model with a
general relativistic treatment of the photon propagation in a Kerr
spacetime surrounding a rapidly rotating neutron star.   This
treatment is essentially valid to first order in the spin of the star
$\Omega$ because it assumes a particular relationship between the
higher moments of the gravitational field that is not valid to second
order.  Furthermore, it neglects the distortion of the stellar surface
due to rapid rotation.  Neglecting the rotational perturbations to the
metric at the beginning of \S~\ref{sec:results} yields an estimate 
of the error in the former approximation.  Understanding latter
approximation would require modelling the structure of the neutron
star in full general relativity \citep[e.g.][]{1995ApJ...444..306S}.
The main thrust of this paper is to confront theoretical predictions
with observations, such as those of
\citet{2002ApJ...581..550M,2003ApJ...595.1066M}, so this latter
comparison is beyond the scope of this paper.

\section{Calculational Overview}
\label{sec:calc-overv}

As \citet{Heyl01typei} assumed, the r-mode perturbs the local
effective temperature of the emission by an amount proportional to the
instantaneous amplitude of the mode at the location.  Furthermore the
emergent spectrum is taken to be a blackbody.  However, unlike
\citet{Heyl01typei}, the calculation here will account for the rapid
rotation of the star and examine the detailed shape of the light
curves, not only the pulsed fraction.  The rapid rotation affects the
observed profiles in three ways.

First, the spacetime surrounding the star no longer is Schwarzschild.
The exterior spacetime will be approximated by the Kerr geometry which
assumes a particular ratio between the higher moments of the
gravitational field.  The distortion of the star due to the rapid
rotation is not included.  The results show that the perturbation to
the geometry is not important at least for stars rotating up to
600~Hz.  \citet{2005ApJ...618..451C} found that a Schwarschild plus 
Doppler treatment (in which the moment of inertia vanishes) gives
good results {\em vis a vis} a fully relativistic treatment.
Second, the photons from different parts of the surface take
different amounts of time to reach the observer.  During this delay
the star may rotate significantly.  Third, there is a significant
difference in the redshift from different regions of the stellar
surface.  To examine all of the these effects separately, the fully
relativistic calculations will be contrasted with rapidly rotating
stars with a vanishing moment of inertia and models where the time
delay and redshift factors are separately neglected.

To determine what observations of the r-mode oscillation may reveal,
several values for the strength of the r-mode and its frequency are
studied.  Also the radius and spin frequency of the star, and the line
of sight of the observer are varied.  Table~\ref{tab:parameters} shows
the various parameters and the values considered.  Although this study
considers only a single value for the stellar mass, results may be
obtained for different masses by scaling the stellar radius with the
mass and the spin frequency inversely with the mass.  If one includes
the models calculated with various physical effects neglected, the
total number is 1200.  The total flux and flux densities at five
different energies are calculated as a function of phase to determine
the pulsed fraction and harmonic content of the oscillations.
\begin{deluxetable}{l|ll}
\tablecaption{Parameter Values
\label{tab:parameters}
}
\tablehead{\colhead{Parameter} & \colhead{Definition} & \colhead{Values} }
\startdata
$M$ & Mass of star & 1.4 M$_\odot$ \\
$f$   & Spin frequency of star & 0, 300, 600, 1200 Hz \\
$i$   & Latitude of line of sight &
5.739$^\circ$, 17.46$^\circ$, 30$^\circ$, 44.43$^\circ$, 64.18$^\circ$
\\
$\sin i$ & Cumulative solid angle & 0.1, 0.3, 0.5, 0.7, 0.9 \\
$R$   & Stellar radius & 8, 10, 12, 14~km\\
$R c^2/(G M)$ & Stellar compactness & 3.87, 4.84, 5.80, 6.77 \\
$h \nu/k T_0$\tablenotemark{a} & Dimensionless photon energy  &
0.3, 1, 3, 10, 30 \\
$A$ & R-mode amplitude & 0.05, 0.15, 0.5 \\
$q\equiv 2\pi f/\omega$ & Spin-mode frequency ratio & 100, 300, 600, 1200 \\ 
\tablenotetext{a}{This is the ratio of the observed energy of the emission to the 
observed blackbody energy of the unpulsed emission.}
\enddata
\end{deluxetable}

The next section presents the details of how the light curves are
calculated (\S~\ref{sec:calc-deta}).  A reader not interested in these
details may be forgiven for skipping to \S~\ref{sec:results} that presents the
results.  \S~\ref{sec:discussion} relates these results both to
the observations and previous theoretical work.

\section{Calculational Details}
\label{sec:calc-deta}

The calculation follows the techniques outlined in the appendix of
\citet{1989ApJ...339..279C}.  Specifically, the photon paths begin at
the observer and are evenly spaced over the image plane.  Before
integrating a particular path, the constants of the photon's motion
are determined: its energy, $E=-P_t$, its specific angular momentum,
$l=-P_\phi/P_t$ and its specific Carter constant $q=P_\theta^2/P_t^2-a^2
\cos^2\theta + l^2 \cot^2 \theta$.    If the value of $q$ is too
small or $l$ is outside a particular range, the photon will not reach
the stellar surface, so its path is not integrated.   Those paths for
those photons that pass these criteria are integrated until either
they reach the surface or they reach their minimum radius.  If the ray
intersects the surface, the redshift of the surface relative to
infinity including the velocity of the surface element is calculated.

\citet{1989ApJ...339..279C} present the variables and metric in
detail, but for clarity the key equations are
\begin{eqnarray}
\dd{\theta}{r} &=& \pm \frac{ (q - l^2 \cot^2 \theta + a^2 \cos^2
\theta)^{1/2} }{V_r^{1/2}} \\
\dd{\phi}{r} &=& \frac{ -a + l \csc^2\theta + a(r^2+a^2-l a)/\Delta }
{V_r^{1/2}} \\
\dd{t}{r} &=& \frac{D - 2 M a l r}{V_r^{1/2} \Delta}
\end{eqnarray}
where 
\begin{eqnarray}
\Delta &=& r^2 - 2 M r + a^2 \\
D &=& (r^2+a^2)^2 - a^2 \Delta \sin^2 \theta \\
V_r &=& (r^2 +a^2 - l a)^2 - \Delta (l^2 + a^2 - 2 l a + q) .
\end{eqnarray}
These equations are expressed in units with $G=c=1$.
Asymptotically far from the star, $r, \theta$ and $\phi$ are the usual
spherical coordinates and $t$ is the time coordinate.  $a$ is the
ratio of the angular momentum of the star to its mass.  The results of
\citet{1994ApJ...424..846R} give $J=0.21 M R^2 \Omega / (1 - 2 G M / R
c^2)$ for slowly rotating neutron stars.   For the models outlined in 
Table~\ref{tab:parameters}, $a/M$ ranges from 0 to $0.721$.

Because the integration begins at the detector and proceeds toward the
star, the rays can be spaced equally in solid angle at infinity.
Because the locally measured intensity divided by fourth power of the
locally measured energy is a relativistic invariant along a ray
bundle, it is straightforward to integrate the total intensity over
the image to obtain the total flux from the surface of the star.  What
remains is to specify the intensity as a function of direction and
position on the surface.  For simplicity, here the radiation is
assumed to be a blackbody, which means that the intensity is
isotropic, and the distribution is specified by a single parameter, the
temperature.  \S~\ref{sec:discussion} outlines some of the
complications of using a more complicated model for the emission.

As \citet{Heyl01typei} assumed, the temperature is proportional to a
uniform component plus a component proportional to the local amplitude
of the r-mode,
\begin{equation}
T(\theta,\phi) = T_0 \left \{  1 +\frac{A}{2} \exp \left (\frac{3}{4} 
- \frac{1}{2} \eta^2 \right ) \left (\frac{1}{2} + \eta^2 \right )
\exp \left [ i \phi - i ( \Omega - \omega  ) t  \right ] \right \}
\label{eq:tperturb}
\end{equation}
where $\eta = \cos\theta \sqrt{q/3}$ and $\Omega = 2 \pi f$.  The
results of \citet{Heyl01typei} and \citet{Long68} have been normalized 
and specialized for an r-mode with $q \gg\ 1$ and $m=\nu=1$.

The specific intensity is taken to be given by the blackbody
formula \citep[e.g][]{Shu91}
\begin{equation}
I_\nu = B_\nu(T) = \frac{2 h \nu^3}{c^2} \frac{1}{e^{h\nu/kT}-1}
\label{eq:specific-intensity}
\end{equation}
and the total intensity is given by
\begin{equation}
\int I_\nu \d \nu = \frac{2 \pi^5}{15} \frac{k^4}{c^2 h^3} T^4.
\label{eq:total-intensity}
\end{equation}
Here only the temperature and frequency dependences are important,
because the calculation is only concerned with relative variations in
the flux not its absolute value.

An examination of Eqs.~\ref{eq:tperturb}-\ref{eq:total-intensity}
shows how the waveform may develop higher harmonics.  If the r-mode
were taken to perturb the intensity directly, there would be no
harmonics.  Because the total observed flux would in this case be
simply a weighed integral of the intensities over the surface of the
star, the varying portion of the flux would be proportional to $\exp
\left [ - i (\Omega -\omega) t \right ]$.

In the case considered here, the r-mode perturbs the temperature, so
harmonics are naturally produced.  The arguments from the preceding
paragraph indicate the fundamental is simply proportional to $A$ and
only three higher harmonics are present in the total flux which are
proportional to $A^2$, $A^3$ and $A^4$ respectively.  To lowest order
in $A$, the pulsed fraction is proportional to $A$, so the ratio of
the higher harmonics to the fundamental decreases rapidly as the
pulsed fraction decreases.  Because the specific intensity is a more
complicated function of temperature (Eq.~\ref{eq:specific-intensity}),
when one looks at the variation at a specific energy, additional
harmonics are generated, but each successive harmonic is weaker by a
factor of $A$.  This should be contrasted with the presence of a hot
region \citep[e.g.][]{2002ApJ...581..550M}.  In the latter case, the
ratio of the higher harmonics to the fundamental is set by the shape
of the hot region and is independent of the amplitude of the
temperature perturbation.

\section{Results}
\label{sec:results}

To examine a wide range of physical effects and geometries, a large
number of light curves were calculated.  To appreciate how the various
physical effects affect the light curves, some sample light curves
will be presented.   A detailed discussion of how the light curves
change as a function of energy, inclination, spin frequency, stellar
compactness and the properties of the r-mode itself follows.

\begin{figure}
\plotone{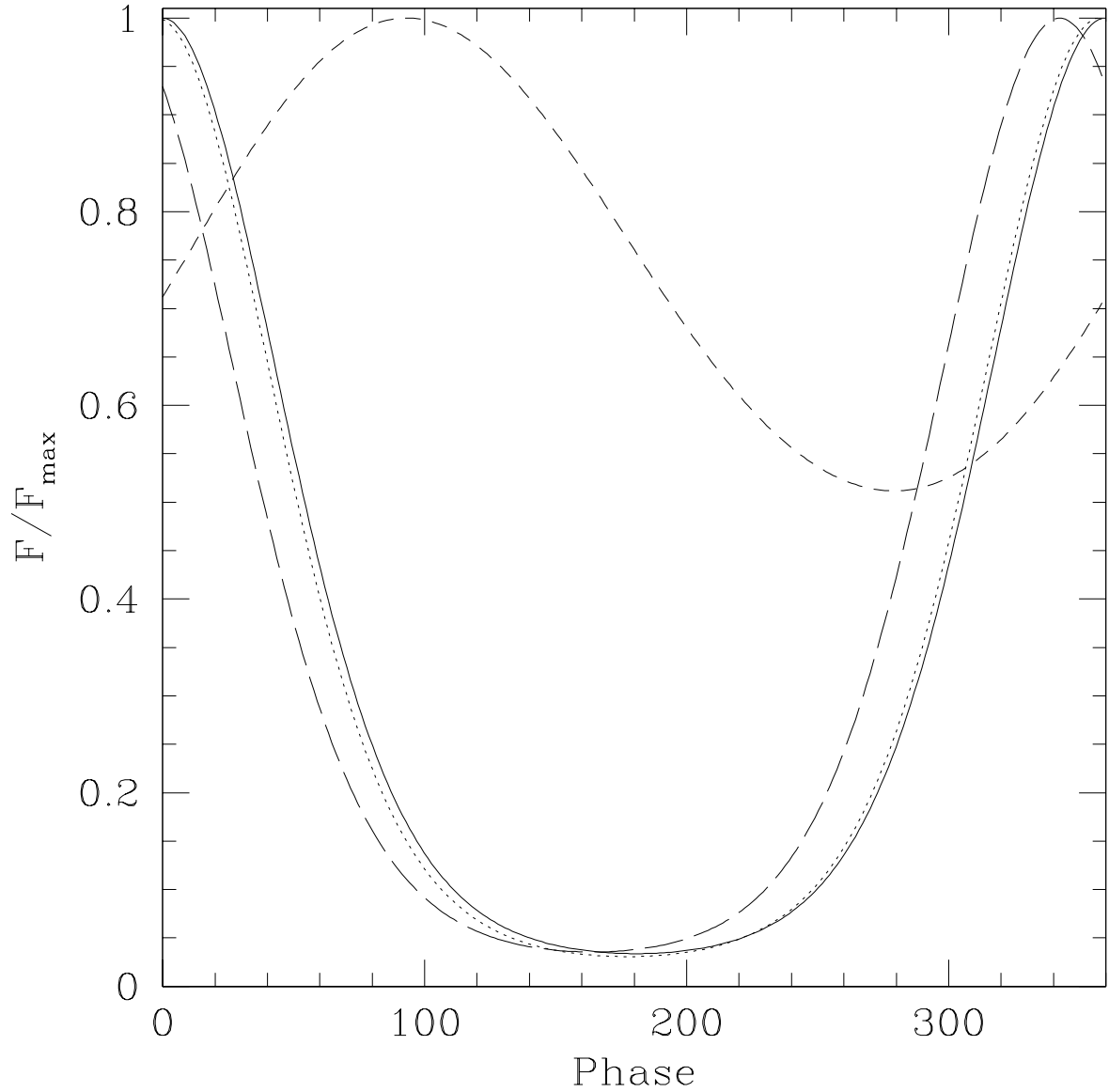}
\caption{The light curves for a rapidly rotating neutron star with
various physical effects neglected.  The solid curve gives the light
curve with all relativistic effects included.  The dotted curve
neglects the perturbation to the spacetime induced by the spin of the
star.  The long-dashed curve neglects the time delay from various
parts of the surface, and the short-dashed curve neglects the
variation in the redshift across the surface of the star.
$f=1200$~Hz, $R/M=3.87$, $i=5.739^\circ$, $A=0.05$ and $q=100$.  
}
\label{fig:physical}
\end{figure}

The effects of the varying redshift on the surface of the star, the
time delay from different regions on the surface, and the inclusion of
the perturbation of the exterior spacetime of the star due to rotation
will be largest for star with the smallest radius and fastest spin
rate viewed from near the equatorial plane.  Fig.~\ref{fig:physical}
depicts the light curves for a neutron star with $f=1200$~Hz,
$R/M=3.87$ and an inclination of 5.739$^\circ$.  

The largest correction is obviously the varying redshift across the
stellar surface.  The varying redshift moves the peak of the emission
earlier in phase, increases the pulsed fraction, and adds harmonics to
the shape of the pulse.  If one neglects frame dragging or varying
time delay to different parts, the difference are much more modest.  The
frame dragging simply delays the phase of the decrease in the emission
by a few degrees.  The phase of the upswing is delayed by a slightly
smaller amount.  The time delay delays the phase of the decrease by
about 15 degrees and delays the upswing by a slightly larger amount.
Because only phase differences are observable, the main consequence of
the time delay and the frame dragging is a very modest change in the
shape of the light curves.

\subsection{Energy Dependence}
\label{sec:energy-dependence}

\begin{figure}
\plottwo{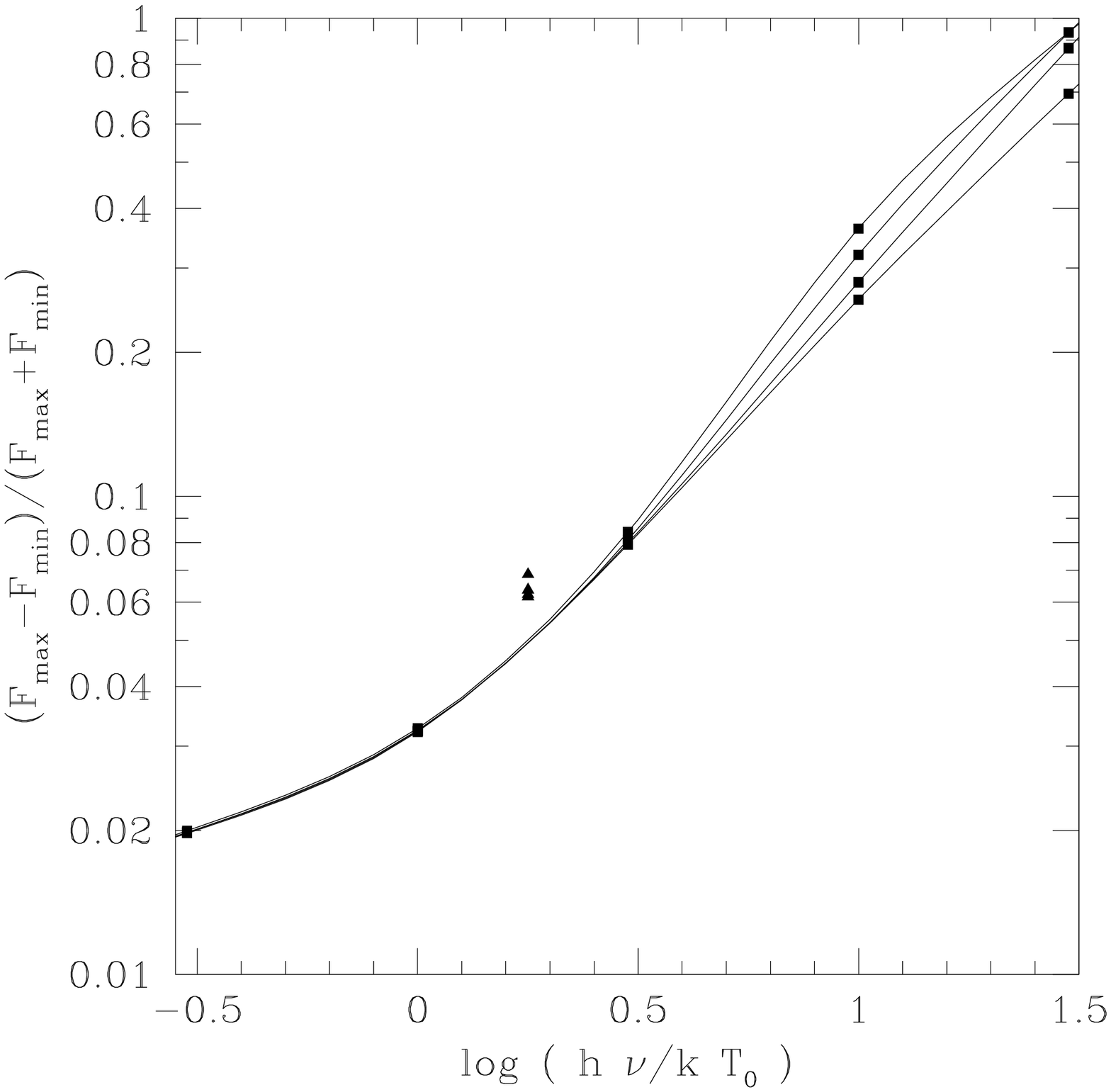}{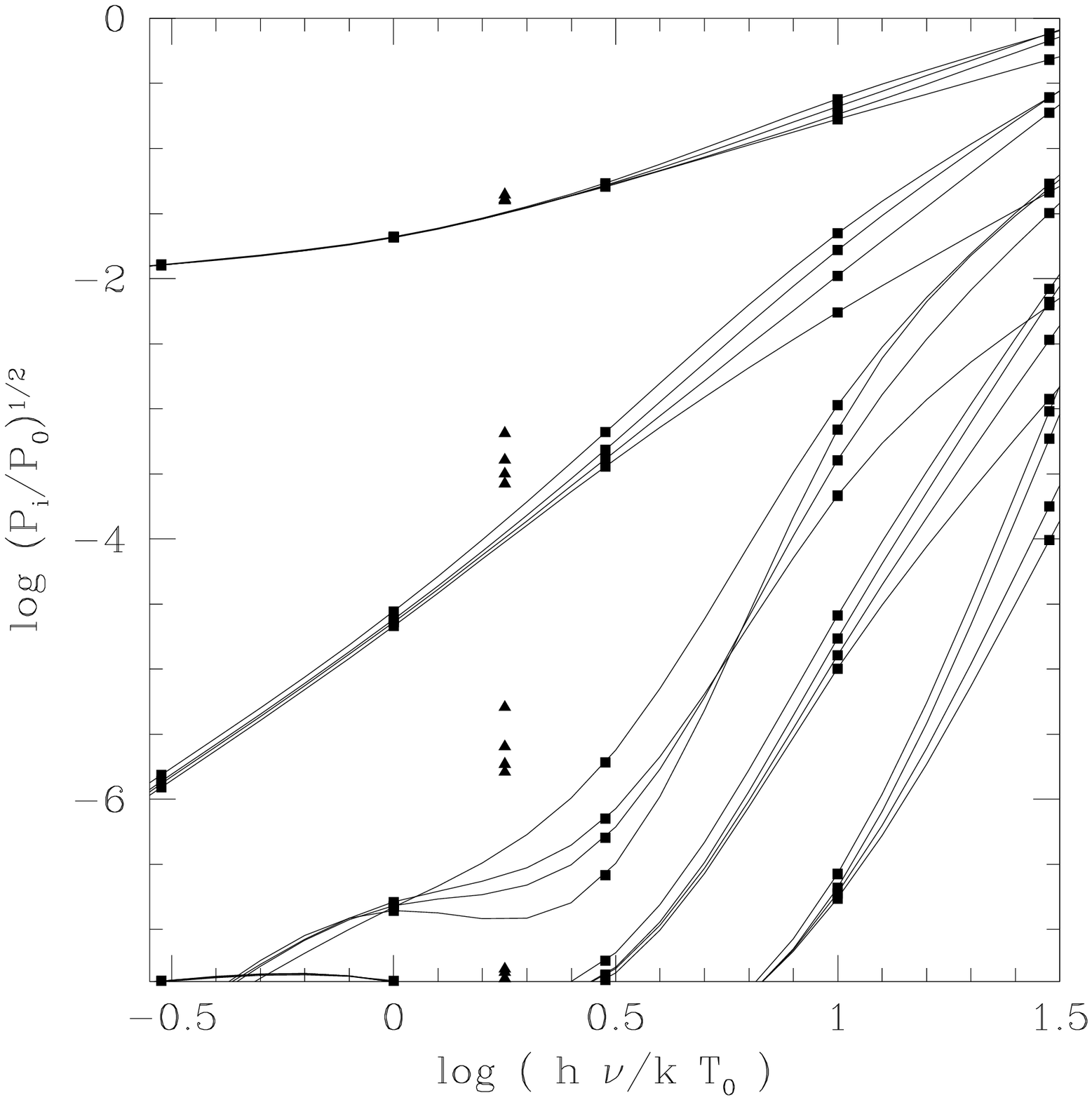}
\caption{The pulsed fraction and harmonic content as a function of energy.
In the left panel, the curves from bottom to top are for stellar spin
frequencies $f=0,300,600$ and $1200$~Hz.  The triangular symbols give 
the result for the total flux.
 In the right panel, each bundle of four curves traces the power
in a successive harmonic, starting at the top with the first harmonic.  
Within each bundle the four curves traces the same stellar spin frequencies 
as in the left panel.  The other parameters are held constant 
at $i=30^\circ$, $R/M=4.84$, $A=0.05$ and $q=100$}
\label{fig:pfe}
\end{figure}

The properties of the burst oscillations dependly strongly on energy
as found by \citet{2003ApJ...595.1066M}.  Specifically, the pulsed
fraction increases dramatically with the energy of the radiation.
Because the pulsed fraction is essentially proportional to the
strength of the first harmonic, and the ratio of two consecutive
harmonics scales with strength of the first harmonic.  The harmonic
content increases dramatically with increasing photon energy as shown
in Fig.~\ref{fig:pfe}.  Because the emission was assumed to have a
blackhody spectrum, the flux from the surface of the star for $h\nu
\gg\ kT$ is extremely sensitive to changes in the temperature on the
surface.  For emission well above the thermal peak, the pulsed
fraction approaches unity and the higher harmonics become nearly as
important as the fundamental.

\begin{figure}
\plotone{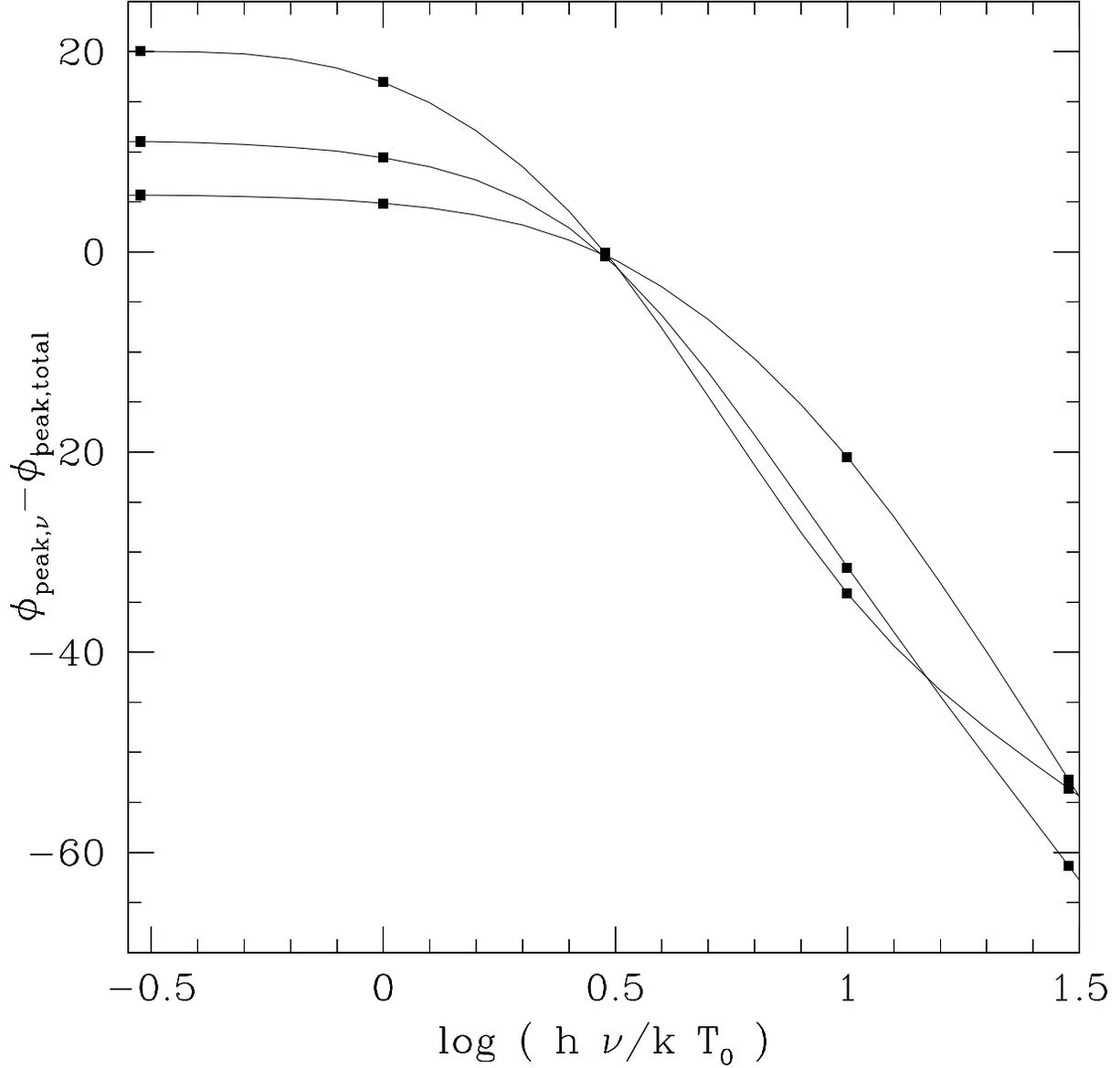}
\caption{Phase lag as a function of photon energy and stellar spin.
The curves depict the phase lag of the peak at a particular energy 
of the emission relative to peak of the total flux.  The peak at
low energies comes after the peak of the total flux, while the peak
at high energies leads the total flux.  The largest absolute values
of the lag/lead are for the fastest spinning stars.  The curves
trace $f=300,600$ and $1200$~Hz and the other parameters
are at $i=30^\circ$, $R/M=4.84$, $A=0.05$ and $q=100$}
\label{fig:leade}
\end{figure}

The Doppler boosting of radiation originating from the half of the
star approaching the observer causes the total flux from the surface
to peak before the hot region crosses the line of the sight to the
observer (see Fig.~\ref{fig:physical}).  As for the pulsed fraction, a
slight increase in the ratio between the observed frequency and
emitted frequency of photons in the Wein tail of the thermal radiation
spectrum can have a dramatic effect on the observed flux, so higher
energy radiation leads the emission at lower energies as shown in
Fig.~\ref{fig:leade} \citep{1999ApJ...519L..73F,2000ApJ...535L.119F}.

\subsection{Inclination and Compactness}
\label{sec:inclination}

\begin{figure}
\plottwo{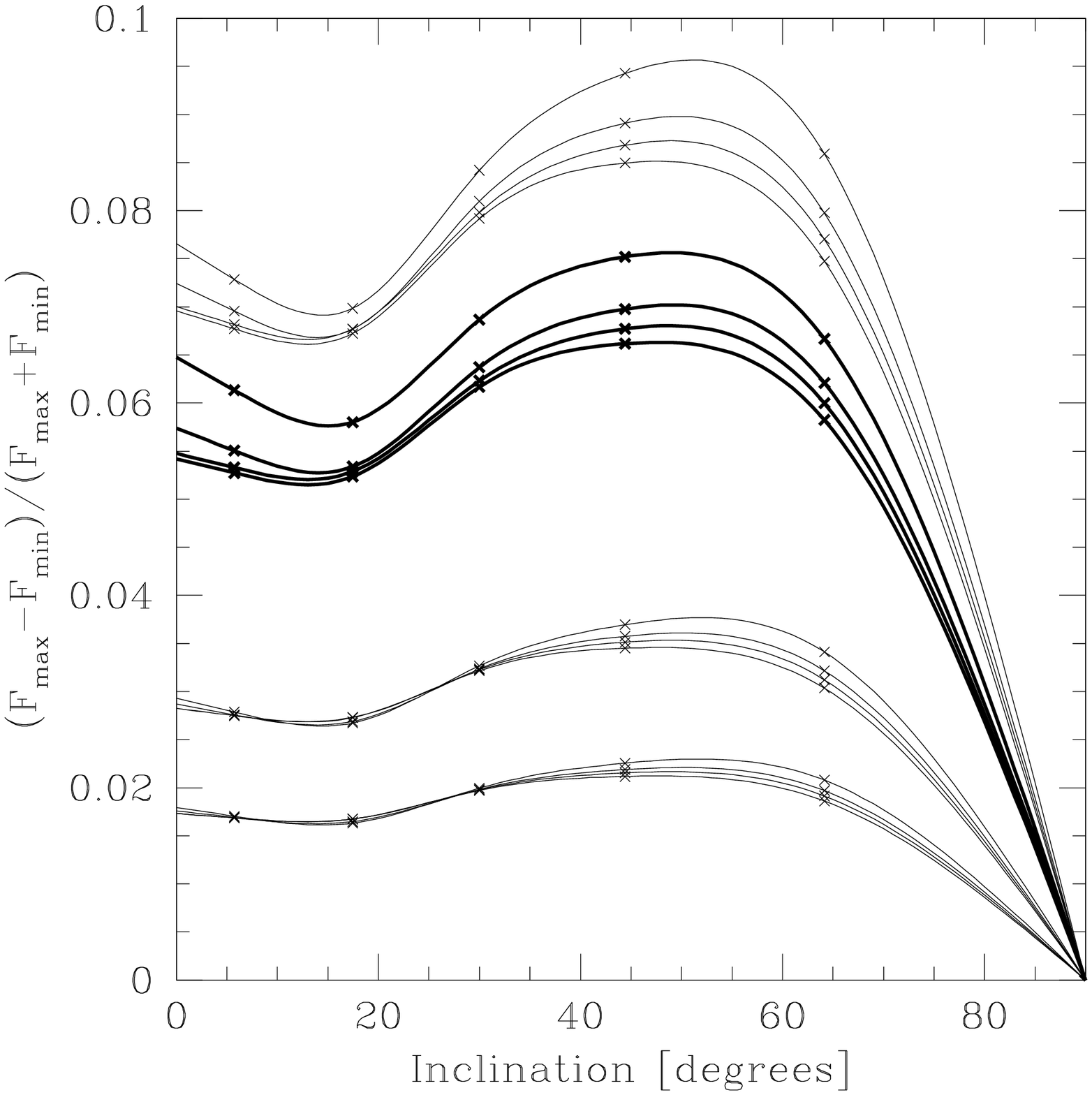}{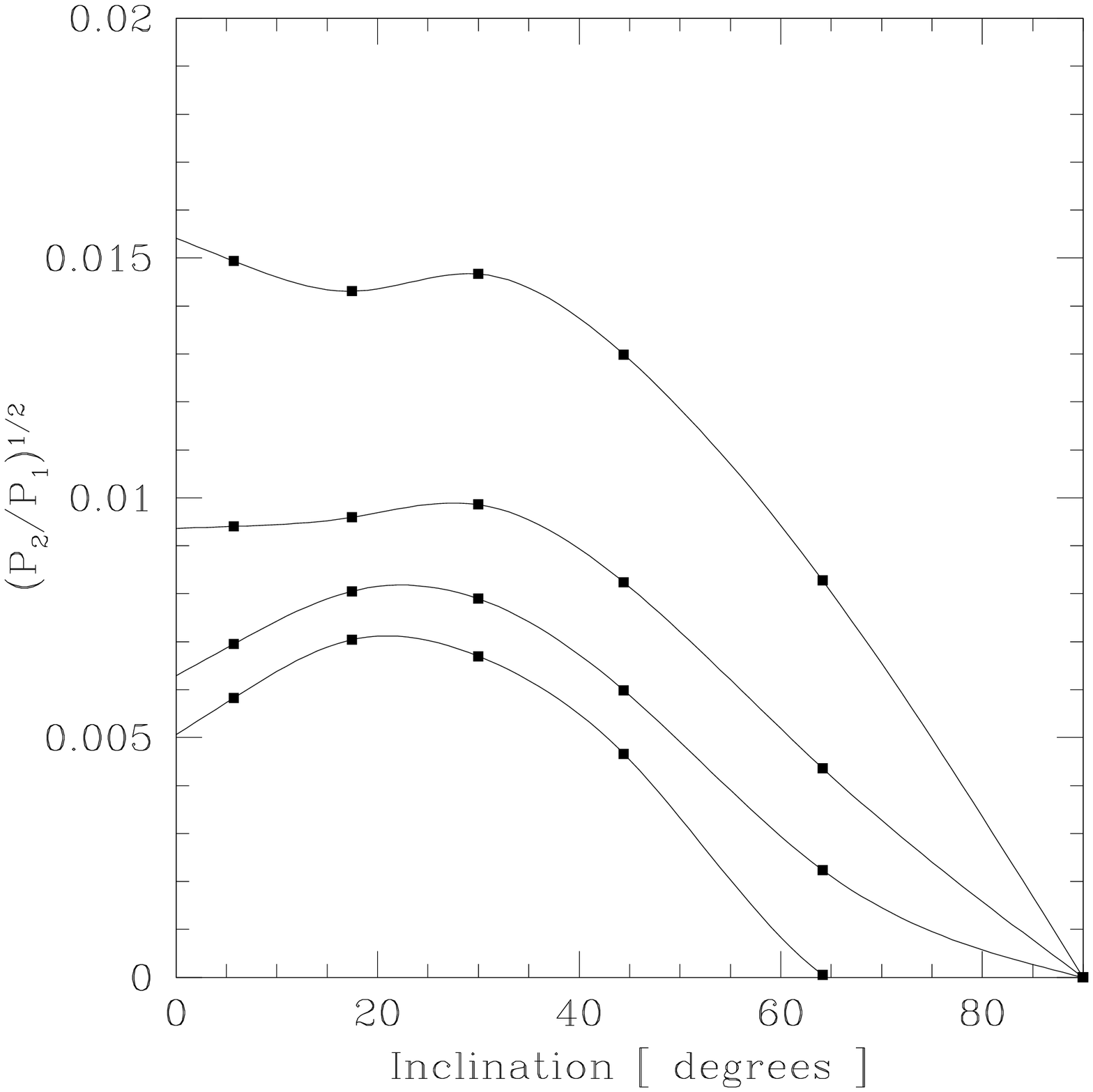}
\caption{The pulsed fraction and harmonic content as a function of inclination.
The the left panel each bundle of four curves depicts the 
results for a different photon energy.  From top to bottom they are .
$h \nu/k T_0=3$, the total flux (bold) and $h \nu/k T_0=1$ and 0.3.
Within each bundle are the results for different spins, 
$f=0,300,600$ and $1200$~Hz from bottom to top.  The right panel
depicts the ratio of the power of the total flux in the second harmonic 
to that in
the first harmonic.  From bottom to top are depicted the different
stellar spins of $f=0,300,600$ and $1200$~Hz.  
The other parameters are fixed at $R/M=4.84$, $A=0.05$ and $q=100$}
\label{fig:pfi}
\end{figure}

The observed light curves of course depend on the inclination of
equator of the star relative to our line of sight and the compactness
of the star, $R/M$.  Because the hot region lies along the equator of
the star, the pulse fraction and harmonic content depend only weakly
on the inclincation of the spin axis of the star (Fig.~\ref{fig:pfi}).
Of course, the pulse fraction and harmonic content vanish for
$i=90^\circ$, but for inclinations up to 64.18$^\circ$, i.e. over 90\%
of the lines of sight, the pulse fraction is nearly constant.

\begin{figure}
\plottwo{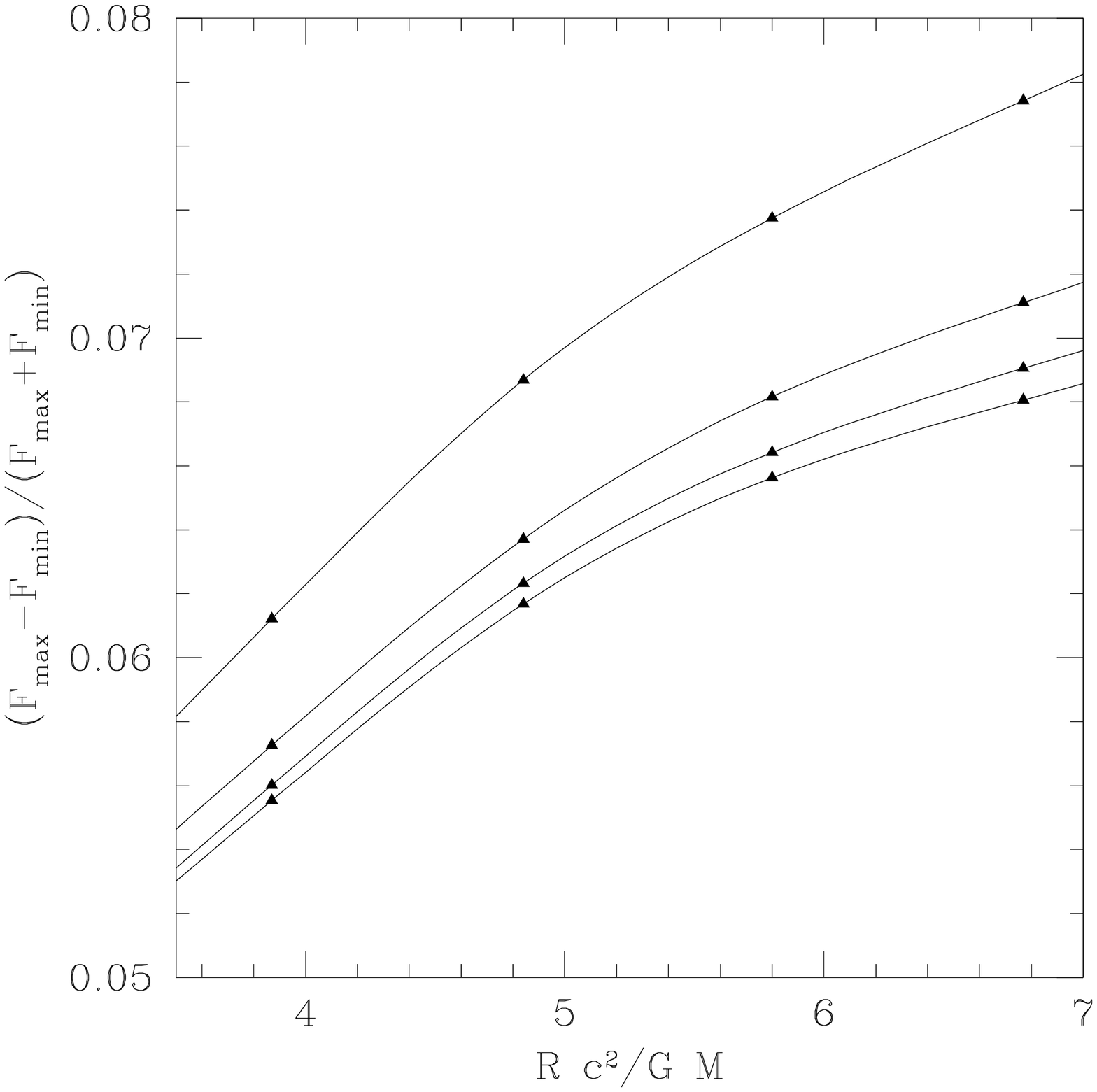}{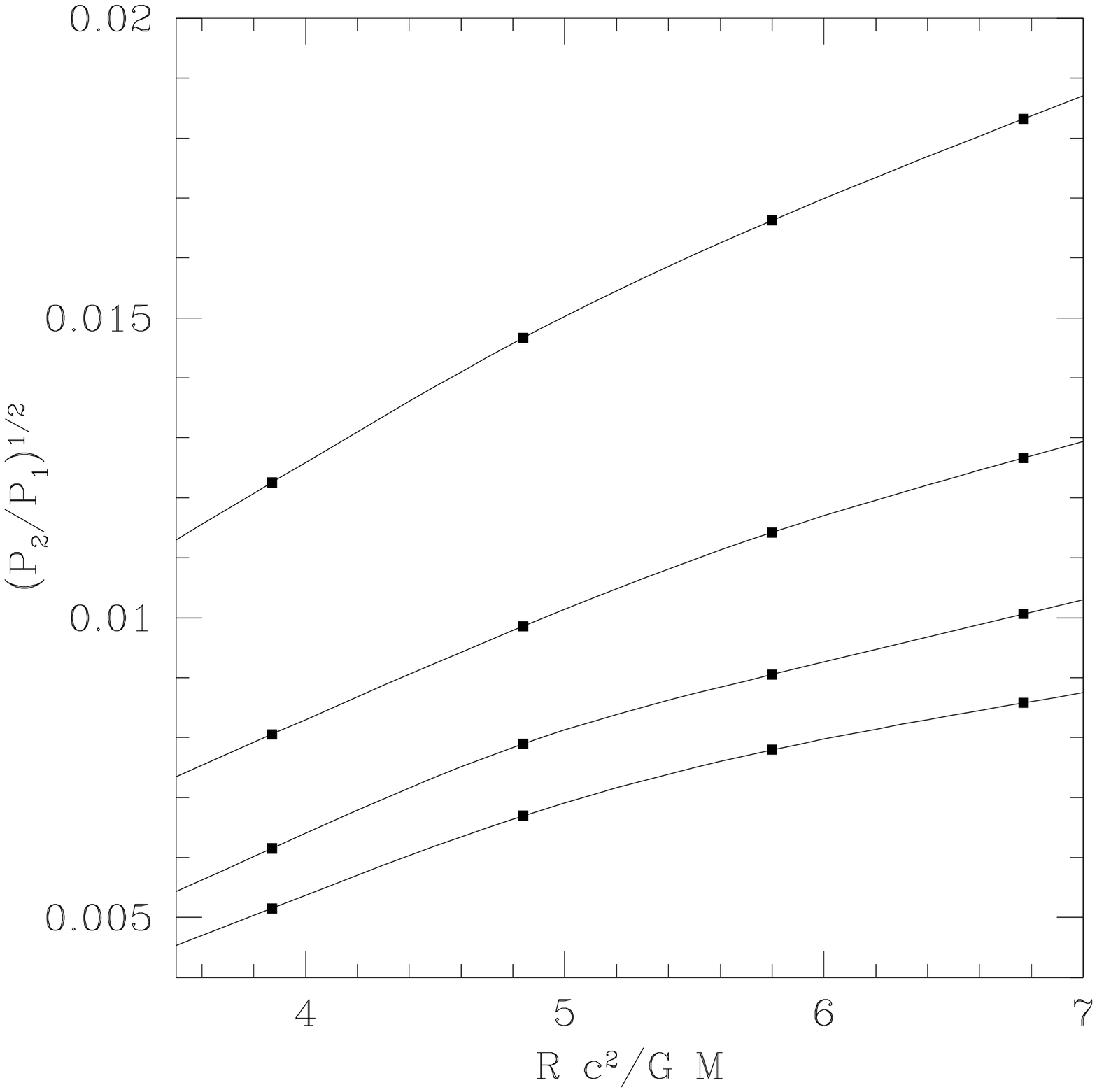}
\caption{The pulsed fraction and harmonic content as a function of radius.
The left panel depicts the total pulsed fraction as a function of stellar
radius and spin.
The right panel depicts the ratio of the power of the total flux in
the second harmonic to that in the first harmonic.  From bottom to top in 
panel the curves are the different values of stellar spin, $f=0,300,600$ and
$1200$~Hz.  The other parameters are fixed at $i=30^\circ$, $A=0.05$
and $q=100$}
\label{fig:pfr}
\end{figure}

As is well known \citep{Page96} more compact stars typically exhibit
less variability because the gravity bends null trajectories from the
rear of the star to reach the observers; therefore, a equatorial hot
region may be visible during an entire rotation of the star.
Fig.~\ref{fig:pfr} bears this out for the r-mode oscillations.  The
r-modes are uniquely sensitive to the physically interesting
compactness while being relatively insensitive to the inclination of
the star \citep[\cf the results of][for hot spots]{2001ApJ...546.1098W,2002ApJ...581..550M}.
This makes them a potentially important probe of the
equation of state of neutron stars.

\subsection{R-mode properties}
\label{sec:r-mode-properties}

To examine how the properties of the r-mode affect the resulting light
curves, both the value of $q = 2\pi f/\omega$ and the amplitude of the
r-mode ($A$) are varied.  As \citet{Heyl01typei} noted as $q$ varies
both the size of the r-mode on the surface of the star changes and the 
observed frequency of the oscillation $f ( 1 - 1/q )$.  Here the focus
will be the former effect exclusively.
\begin{figure}
\plotone{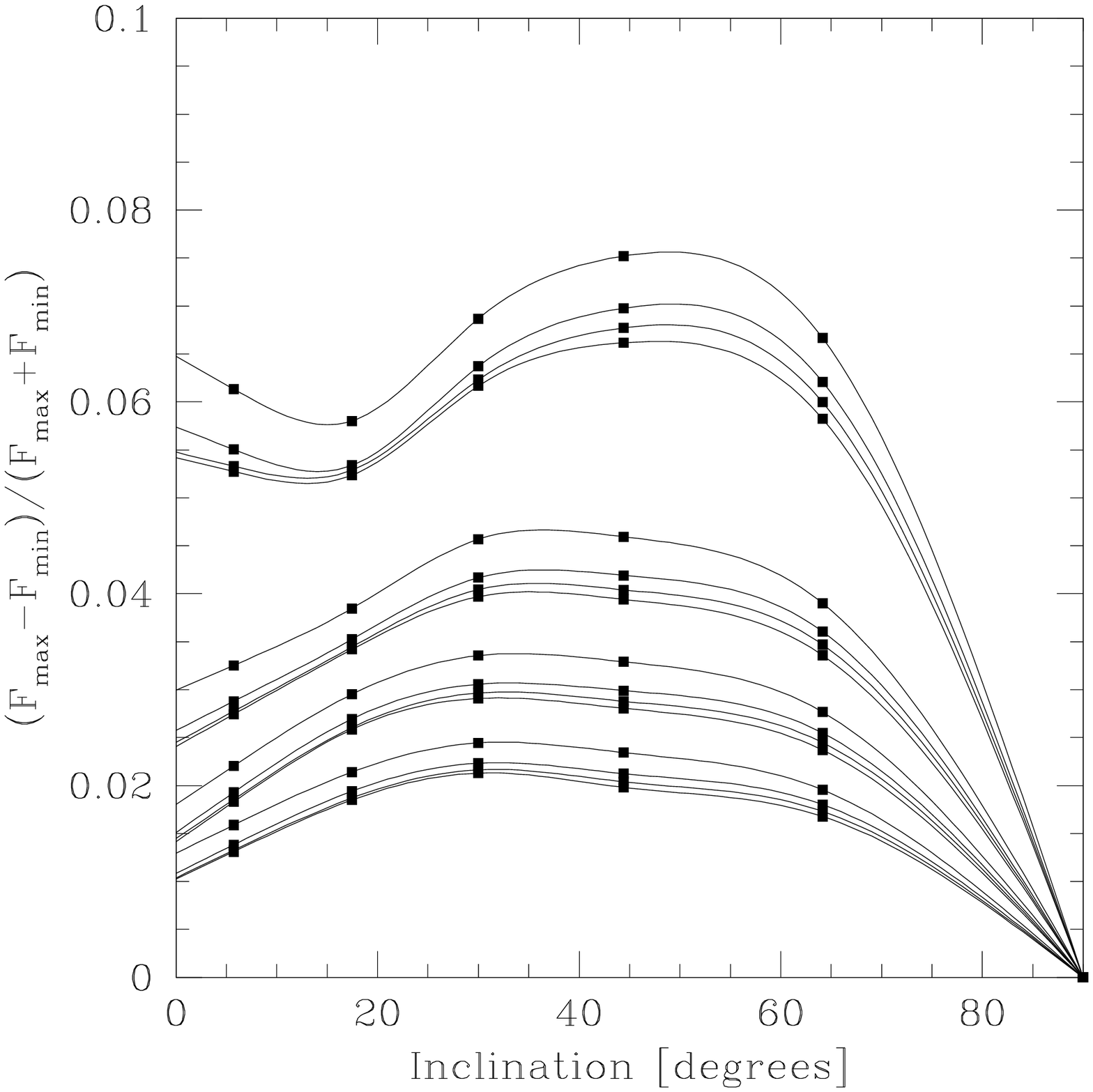}
\caption{Pulsed fraction as a function of $q$.  The curve depicts the
total pulsed fraction as a function of inclination, $q$ and $f$.  The
bundles of curves are from top to bottom are for $q=100,300,600$ and
1200.  Within each bundle are the stellar spins from bottom to top of
$f=0,300,600$ and $1200$~Hz.}
\label{fig:pfqq}
\end{figure}

\begin{figure}
\plotone{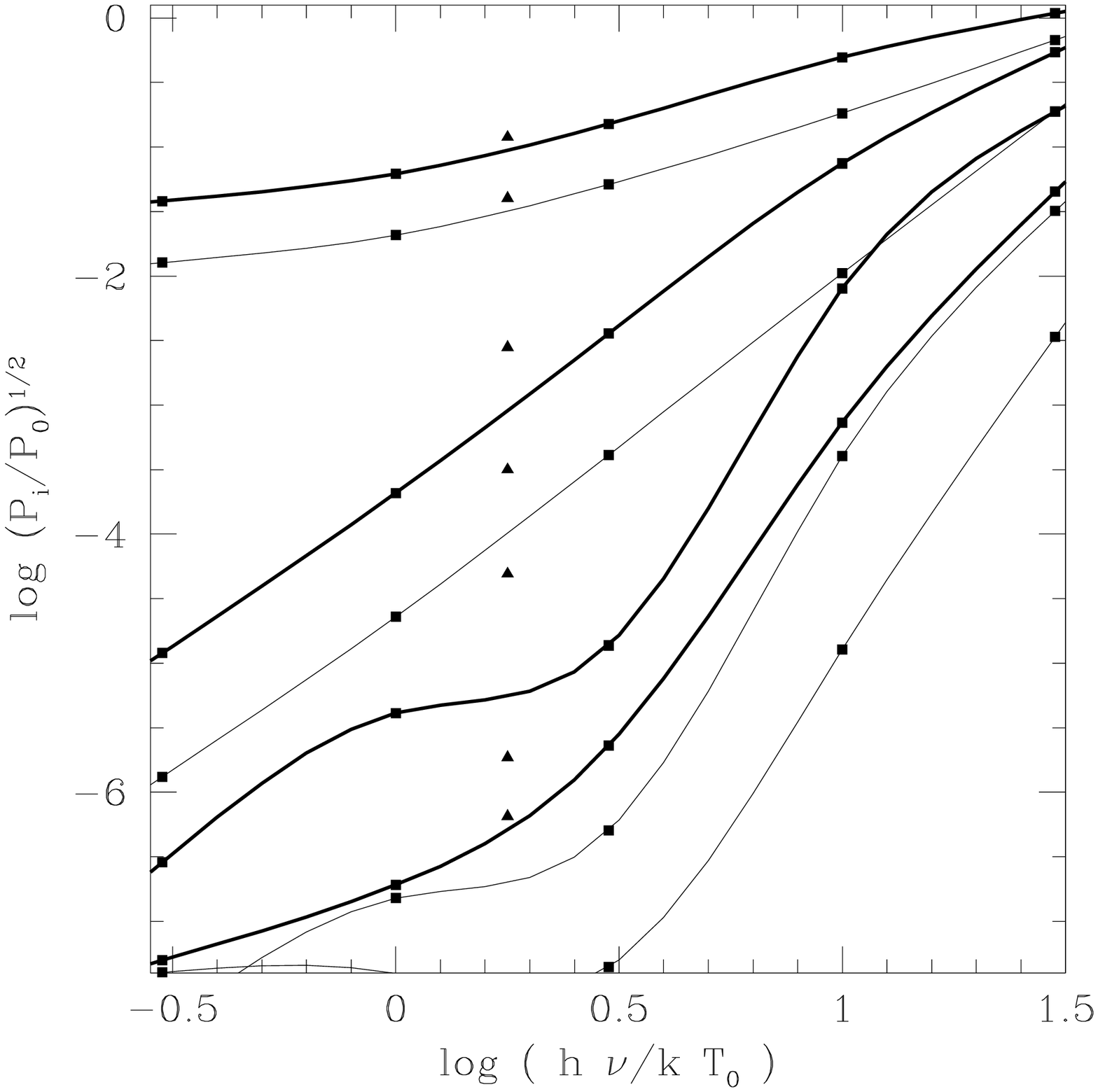}
\caption{Harmonic content as a function of $A$.  The bold curves have
$A=0.15$ and the light curves have $A=0.05$. The triangles give the
result for the total flux, and the curves trace the ratio of the flux
in the first, second, third and fourth harmonics from top to bottom.
The other parameters are $i=30^\circ$, $R=4.84 M=10$~km, $q=100$ and
$f=300$~Hz.}
\label{fig:harmrr}
\end{figure}

The end of \S~\ref{sec:calc-deta} argues that the ratio of the $i$
harmonic to the $(i+1)$ harmonic is simply $A$.  Fig.~\ref{fig:harmrr}
demonstrates that this is indeed the case even as a function of the
photon energy.  This result contrasts with the results for a hot spot
that exhibits a constant ratio between the strengths of the various
harmonics.  This effect is a robust prediction of the r-mode model for
Type-I X-ray burst oscillations.

\section{Discussion}
\label{sec:discussion}

The paper examines the properties of the variability in the flux from
the surfaces of neutron stars from Type-I X-ray burst oscillations.
Because the underlying emission is known and fixed by the detection of
the oscillations themselves, r-modes provide a unique and powerful
tools to understand the properties of the underlying neutron star.
This situation must be constrasted with hot spot models
\citep[\eg][]{2002ApJ...581..550M} in which the size, position and
number of hot spots are {\em a priori} unknown and the observed
lightcurves depend sensitively on these properties.

\citet{2002ApJ...581..550M} found that the amplitude of the
fundamental oscillation during the tail of the x-ray bursts typically
was five to ten percent.  They only obtained upper limits to the
strength of the higher harmonics that were less than a tenth of that
of the fundamental.  These results agree quite nicely with the results
presented in this paper.  This low harmonic content is difficult to
account for in a hot spot model \citep[see Fig.~3--8
  of][]{2002ApJ...581..550M}.

This paper argues that the pulsed fraction should increase
dramatically with increasing photon energy.
\citet{2003ApJ...595.1066M} found just this trend in the burst
oscillations.  However, \citet{2003ApJ...595.1066M} found that the
peak in the emission at high energies lags that at lower energies.
This contradicts the theoretical models presented here as well as those
of \citet{2003ApJ...595.1066M} and \citet{1999ApJ...519L..73F}.
\citet{2003ApJ...595.1066M} argue that the high-energy photons may be
generated through Compton scattering of lower energy photons in the
accretion disk corona \citep[\eg][]{1995ApJ...441..770M}.  A similar
conclusion must be drawn in the context of this model, reducing its
utility to understand the observed oscillations at the highest
energies.

\section{Conclusion}
\label{sec:conclusion}

R-modes on the surface of neutron stars during Type-I X-ray bursts
present a rich phenomenology and may provide a useful new method to
determine the parameters of neutron stars such as their masses and
radii.  When one observes a Type-I X-ray burst oscillation, several of
the parameters listed in Tab.~\ref{tab:parameters} can be obtained
easily -- $f$, the spin frequency of the star and $q$, the ratio of
the spin frequency to the frequency of the mode.  The remaining
parameters, including the mass and radius of the star can be
determined from observations of the pulsed fraction and harmonic
content as a function of energy.

Only two parameters describe the r-mode excitation on the surface of
the star -- $q$ that is determined by measuring the frequency drift of
the burst oscillation \citep{Heyl01typei} and $A$ by calculating the
ratio of the first harmonic to the fundamental frequency of the
oscillation observed at a particular frequency (\eg
Fig.~\ref{fig:harmrr}).  With the properties of the surface pattern
fixed, the pulsed fraction and phase lag as a function of energy can
determine the remaining physical parameters, $R$ and $M$.  The pulse
fraction is not particularly sensitive to the geometric parameter $i$.
Burst oscillations may be a powerful probe of both the physics of
nuclear burning on the surfaces of neutron stars and of the nuclear
equation of state.  To realize the full potential of these techniques
requires relativistic calculations of both the interior of rotating
neutron stars and ray tracing through the external spacetime as well
as atmospheric modelling.

\acknowledgments 
Support for this work was provided by the National Aeronautics and
Spcae Administration through Chandra Postdoctoral Fellowship Award
Number PF0-10015 issued by the Chandra X-ray Observatory Center, which
is operated by the Smithsonian Astrophysical Observatory for and on
behalf of NASA under contract NAS8-39073.

\def\mn{MNRAS}

\bibliographystyle{apj}
\bibliography{mine,physics,ns,gr,typei}

\begin{thebibliography}{14}
\expandafter\ifx\csname natexlab\endcsname\relax\def\natexlab#1{#1}\fi

\bibitem[{{Cadeau} {et~al.}(2005){Cadeau}, {Leahy}, \&
  {Morsink}}]{2005ApJ...618..451C}
{Cadeau}, C., {Leahy}, D.~A., \& {Morsink}, S.~M. 2005, \apj, 618, 451

\bibitem[{{Chen} \& {Shaham}(1989)}]{1989ApJ...339..279C}
{Chen}, K. \& {Shaham}, J. 1989, \apj, 339, 279

\bibitem[{{Ford}(1999)}]{1999ApJ...519L..73F}
{Ford}, E.~C. 1999, \apjl, 519, L73

\bibitem[{{Ford}(2000)}]{2000ApJ...535L.119F}
---. 2000, \apjl, 535, L119

\bibitem[{Heyl(2004)}]{Heyl01typei}
Heyl, J.~S. 2004, \apj, 600, 939

\bibitem[{Longuet-Higgins(1968)}]{Long68}
Longuet-Higgins, M.~S. 1968, Phil. Trans. R. Soc., 262, 511, (LH68)

\bibitem[{{Miller}(1995)}]{1995ApJ...441..770M}
{Miller}, M.~C. 1995, \apj, 441, 770

\bibitem[{{Muno} {et~al.}(2002){Muno}, {{\" O}zel}, \&
  {Chakrabarty}}]{2002ApJ...581..550M}
{Muno}, M.~P., {{\" O}zel}, F., \& {Chakrabarty}, D. 2002, \apj, 581, 550

\bibitem[{{Muno} {et~al.}(2003){Muno}, {{\" O}zel}, \&
  {Chakrabarty}}]{2003ApJ...595.1066M}
---. 2003, \apj, 595, 1066

\bibitem[{Page \& Sarmiento(1996)}]{Page96}
Page, D. \& Sarmiento, A. 1996, ApJ, 473, 1067

\bibitem[{{Ravenhall} \& {Pethick}(1994)}]{1994ApJ...424..846R}
{Ravenhall}, D.~G. \& {Pethick}, C.~J. 1994, \apj, 424, 846

\bibitem[{Shu(1991)}]{Shu91}
Shu, F.~H. 1991, The Physics of Astrophysics: Volume I. Radiation (Mill Valley,
  California: University Science Books)

\bibitem[{{Stergioulas} \& {Friedman}(1995)}]{1995ApJ...444..306S}
{Stergioulas}, N. \& {Friedman}, J.~L. 1995, \apj, 444, 306

\bibitem[{{Weinberg} {et~al.}(2001){Weinberg}, {Miller}, \&
  {Lamb}}]{2001ApJ...546.1098W}
{Weinberg}, N., {Miller}, M.~C., \& {Lamb}, D.~Q. 2001, \apj, 546, 1098

\end{thebibliography}

\end{document}